\documentclass[twocolumn,prl,superscriptaddress,floatfix,showpacs]{revtex4-1}
\usepackage[utf8]{inputenc}
\usepackage{amsmath}
\usepackage{amssymb}
\usepackage{graphicx}

\makeatletter
\usepackage{graphics}
\usepackage{epsfig}
\usepackage{color}
\newcommand{\red}{\textcolor{black}}
\newcommand{\be}{\begin{equation}} \newcommand{\ee}{\end{equation}}
\newcommand{\bea}{\begin{eqnarray}} \newcommand{\eea}{\end{eqnarray}}

\makeatother

\begin{document}

\title{Transition from asynchronous to oscillatory dynamics in balanced spiking networks
with instantaneous synapses}

\author{Matteo di Volo} 
\affiliation{Unit\'e de Neuroscience, Information et Complexit\'e (UNIC),
CNRS FRE 3693, 1 avenue de la Terrasse, 91198 Gif sur Yvette, France}
\author{Alessandro Torcini} 
\affiliation{Laboratoire de Physique Th\'eorique et Mod\'elisation, Universit\'e de Cergy-Pontoise, CNRS, UMR 8089,
95302 Cergy-Pontoise cedex, France}
\affiliation{Max Planck Institut f\"ur Physik komplexer Systeme, N\"othnitzer Str. 38, 01187 Dresden, Germany}
\affiliation{CNR - Consiglio Nazionale delle Ricerche - Istituto dei Sistemi Complessi, via Madonna del Piano 10, 50019 Sesto Fiorentino, Italy}

\date{\today}

\begin{abstract}
We report a transition from asynchronous to oscillatory behaviour in balanced inhibitory networks for class I and II neurons with instantaneous synapses. Collective oscillations emerge for sufficiently connected networks. Their origin is understood in terms of a recently developed mean-field model, whose stable solution is a focus. Microscopic irregular firings, due to balance, trigger sustained oscillations by exciting the relaxation dynamics towards the macroscopic focus.
The same mechanism induces in balanced excitatory-inhibitory networks
quasi-periodic collective oscillations.
\end{abstract}

\maketitle

{\it Introduction.} Cortical neurons fire quite irregularly
and with low firing rates, despite being subject to a continuous bombardment
from thousands of pre-synaptic excitatory and inhibitory neurons \cite{destexhe1999}. 
This apparent paradox can be solved by introducing the concept of balanced network, where excitation and inhibition balance each other and the neurons are kept near their firing threshold \cite{vogels2005}.
In this regime spikes, representing the elementary units of information in the brain,
are elicited by stochastic fluctuations in the net input current yielding an irregular microscopic activity, while neurons can promptly respond to input modifications \cite{deneve2016}.

In neural network models balance can emerge spontaneously in
coupled excitatory and inhibitory populations
thanks to the dynamical adjustment of their firing rates \cite{bal1,brunel2000,bal2,bal3,bal4,wolf}.
The usually observed dynamics is an asynchronous state characterized
by irregular neural firing joined to stationary firing rates \cite{bal1,bal2,bal3,wolf}.
The asynchronous state has been experimentally observed  both in vivo and in vitro \cite{barral2016,dehghani2016dynamic}, however this is not the only state
observable during spontaneous cortical activity.  In particular,
during spontaneous cortical oscillations excitation and inhibition wax and wane together \cite{okun2008}, suggesting that balancing is crucial for the occurrence of these oscillations with inhibition representing the essential component for the emergence of the synchronous activity \cite{isaacson2011,le2016}. 

The emergence of collective oscillations (COs) in inhibitory networks has been widely investigated in networks of spiking leaky integrate-and-fire (LIF) neurons.
In particular, it has been demonstrated that COs emerge from asynchronous states via Hopf bifurcations in presence of an additional time scale, beyond the one associated to the membrane potential evolution, which can be the transmission delay \cite{brunel1999, brunel2000} or a finite synaptic time \cite{van1994}.
As the frequency of the COs is related to such external time scale this mechanism is normally related to fast ($>$30 Hz) oscillations. 
Nevertheless, despite many theoretical studies, it remains unclear which other mechanisms could be invoked to justify the broad range of COs' frequencies observed experimentally \cite{chen2017distinct}.

In this Letter we present a novel mechanism for the emergence of
COs in balanced spiking inhibitory networks in absence of any synaptic or 
delay time scale. In particular, we show for class I and II neurons \cite{izhikevich2007} that COs arise from an asynchronous
state by increasing the network connectivity (in-degree). 
Furthermore, we show that the COs can survive only in presence of
irregular spiking dynamics due to the dynamical balance.
The origin of COs can be explained by considering the phenomenon at
a macroscopic level, in particular we extend an exact mean-field
formulation for the spiking dynamics of Quadratic Integrate-and-Fire (QIF) neurons~\cite{montbrio2015} to sparse balanced networks.
An analytic stability analysis of the mean-field model reveals
that the asymptotic solution for the macroscopic model is a stable focus
and determines the frequency of the associated relaxation oscillations. 
\red{The agreement of this relaxation frequency with the COs' one measured 
in the spiking network suggests that the irregular microscopic firings of the neurons
are responsible for the emergence of sustained COs corresponding to the relaxation dynamics towards the macroscopic focus.}
This mechanism elicits COs through the excitation of an internal macroscopic time scale, that can range from seconds to tens of milliseconds, yielding a broad range of collective oscillatory frequencies. We then analyse balanced excitatory-inhibitory populations revealing the existence of COs characterized by two \red{distinct} frequencies, whose 
emergence is due, also in this case, to the excitation of a mean-field focus induced by fluctuation-driven microscopic dynamics.

{\it The model.}
We consider a balanced network of $N$ pulse-coupled inhibitory neurons, whose membrane potential evolves as
\begin{equation}
\tau_{m}\dot v_i = F(v_i) +I-2 \tau_m g\sum_{j\in pre(i)}\varepsilon_{ij}\delta(t-t_j)
\label{network}
\end{equation}
where $I$ is the external DC current, $g$ is the inhibitory synaptic coupling, $\tau_m=20$ ms is the membrane time constant and fast synapses (idealized as $\delta$-pulses) are considered.
The neurons are randomly connected, with in-degrees $k_i$ distributed according to a Lorentzian PDF peaked at $K$ and with a half-width half-maximum (HWHM) $\Delta_K$.
The elements of the corresponding adjacency matrix $\varepsilon_{ij}$ are one (zero) if the neuron $j$ is connected (or not) to neuron $i$. We consider two paradigmatic models of spiking neuron: 
the quadratic-integrate and fire (QIF) with $F(v) = v^2$ \cite{ermentrout1986},
which is a current-based model of class I excitability; and the Morris-Lecar (ML) \cite{morris1981,model} representing a conductance-based class II excitable membrane.
The DC current and the coupling are rescaled with the median in-degree 
as $I=\sqrt{K} I_0$ and $g=g_0/\sqrt{K}$, as usually done in order to 
achieve a self-sustained balanced state for sufficiently large in-degrees \cite{bal1,bal2,bal3,bal4,jahnke2009,wolf,montefortePRX}. 
Furthermore, in analogy with Erd{\"o}s-Renyi networks 
we assume $\Delta_K=\Delta_0 \sqrt{K}$. We have verified that the 
reported phenomena are not related to the 
peculiar choice of the distribution of the in-degrees, 
namely Lorentzian, needed to obtain an exact 
mean-field formulation for the network evolution \cite{montbrio2015}, 
but that they can be observed also for more standard distributions,
\red{like Erd\"os-Renyi and Gaussian ones (for more details see the SM~\cite{model} and \cite{tobe}).}

In order to characterize the network dynamics we measure the mean membrane potential ${V}(t) = \sum_{i=1}^N v_i(t)/N$, the instantaneous firing rate $R(t)$, corresponding to the number
of spikes emitted per unit of time, as well as 
the population averaged coefficient of variation ${CV}$ \cite{CV}
measuring the fluctuations in the neuron dynamics. Furthermore, the level of 
coherence in the neural activity
can be quantified in terms of the following indicator \cite{scholarpedia}
\begin{equation}
\rho \equiv \left(\frac{\sigma_V^2}
    {\sum_{i=1}^N \sigma^2_i/N}\right)^{1/2} \; ,
\end{equation} 
where $\sigma_{V}$ is the standard deviation
of the mean membrane potential,  $\sigma^2_i = {\langle V_i^2 \rangle}- {\langle V_i \rangle}^2$
and $\langle \cdot \rangle$ denotes a time average. 
A coherent macroscopic activity is associated to a finite value of
$\rho$ (perfect synchrony corresponds to $\rho=1$), 
while an asynchronous dynamics to a vanishingly small
$\rho \approx {\cal O}(1/\sqrt{N})$. Time averages and fluctuations are usually
estimated on time intervals $ \simeq 120$ s, after discarding transients $\simeq 2$ s.

\begin{figure}
\begin{centering}
\includegraphics[width=0.5\textwidth,clip=true]{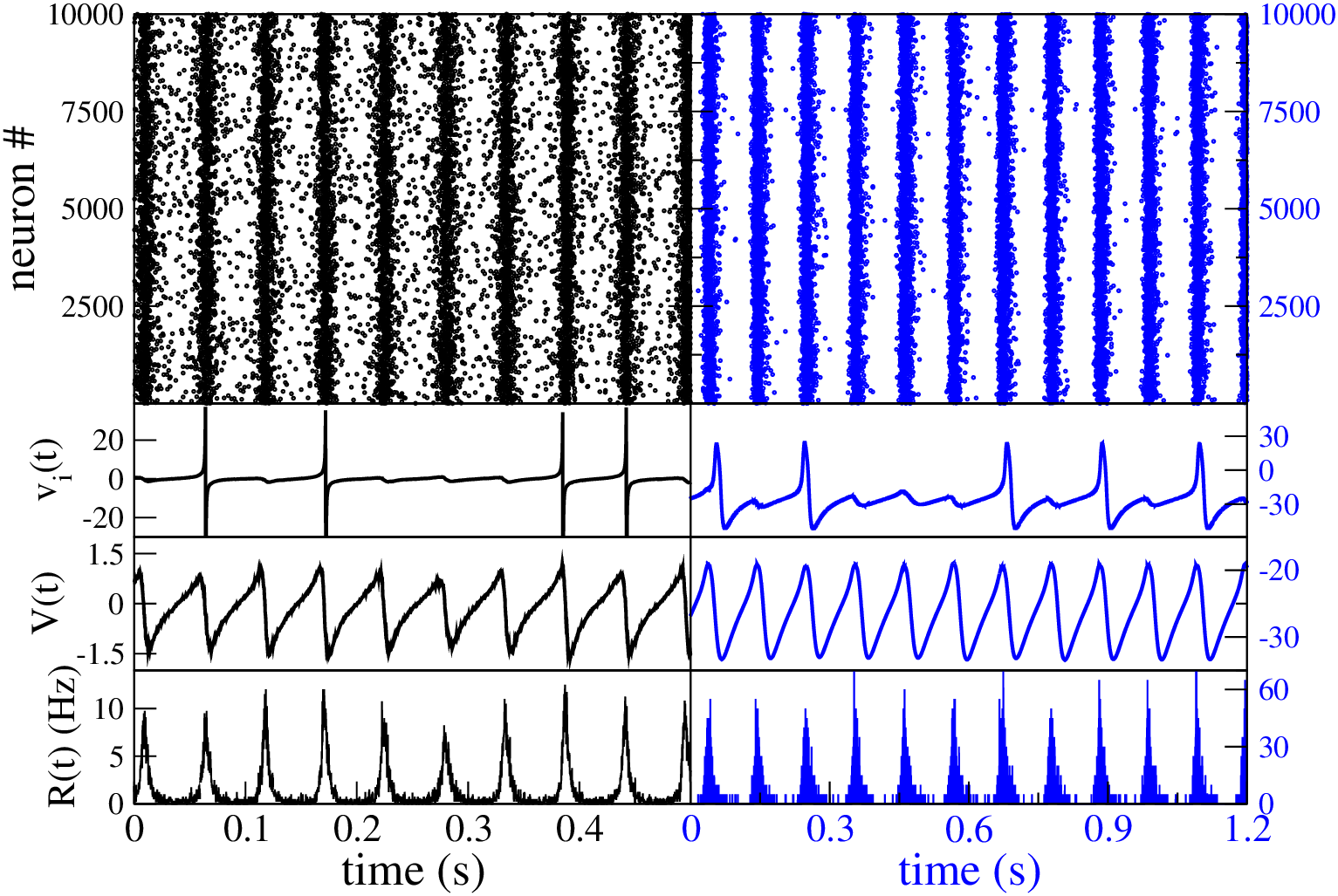}
\end{centering}
\caption{\label{fig:1} The panels show (from top to bottom) the raster plots and the corresponding time traces for the membrane potential $v_i(t)$ of a representative neuron, for ${V}(t)$ and $R(t)$. 
Left row (black): QIF and right row (blue): ML. The parameter values are 
$N=10000$,  $K=1000$, $\Delta=0.3$, $g_0=1$  and $I_0=0.015$.}
\end{figure}

{\it Results.}
In both models we can observe collective firings, or population bursts,
occurring at almost constant frequency $\nu_{osc}$. 
As shown in Fig. \ref{fig:1}, despite the almost regular macroscopic oscillations in
the firing rate $R(t)$ and in the mean membrane potential $V(t)$, the microscopic dynamics 
of the neurons $v_i(t)$ is definitely irregular.
The latter behaviour is expected for balanced networks, where the dynamics 
of the neurons driven by the fluctuations in the input current, 
however usually the collective dynamics is
asynchronous and not characterized by COs as in the present case \cite{bal1,bal2,bal3,bal4,jahnke2009,wolf,montefortePRX}.

\begin{figure}
\begin{centering}
\includegraphics[width=0.5\textwidth,clip=true]{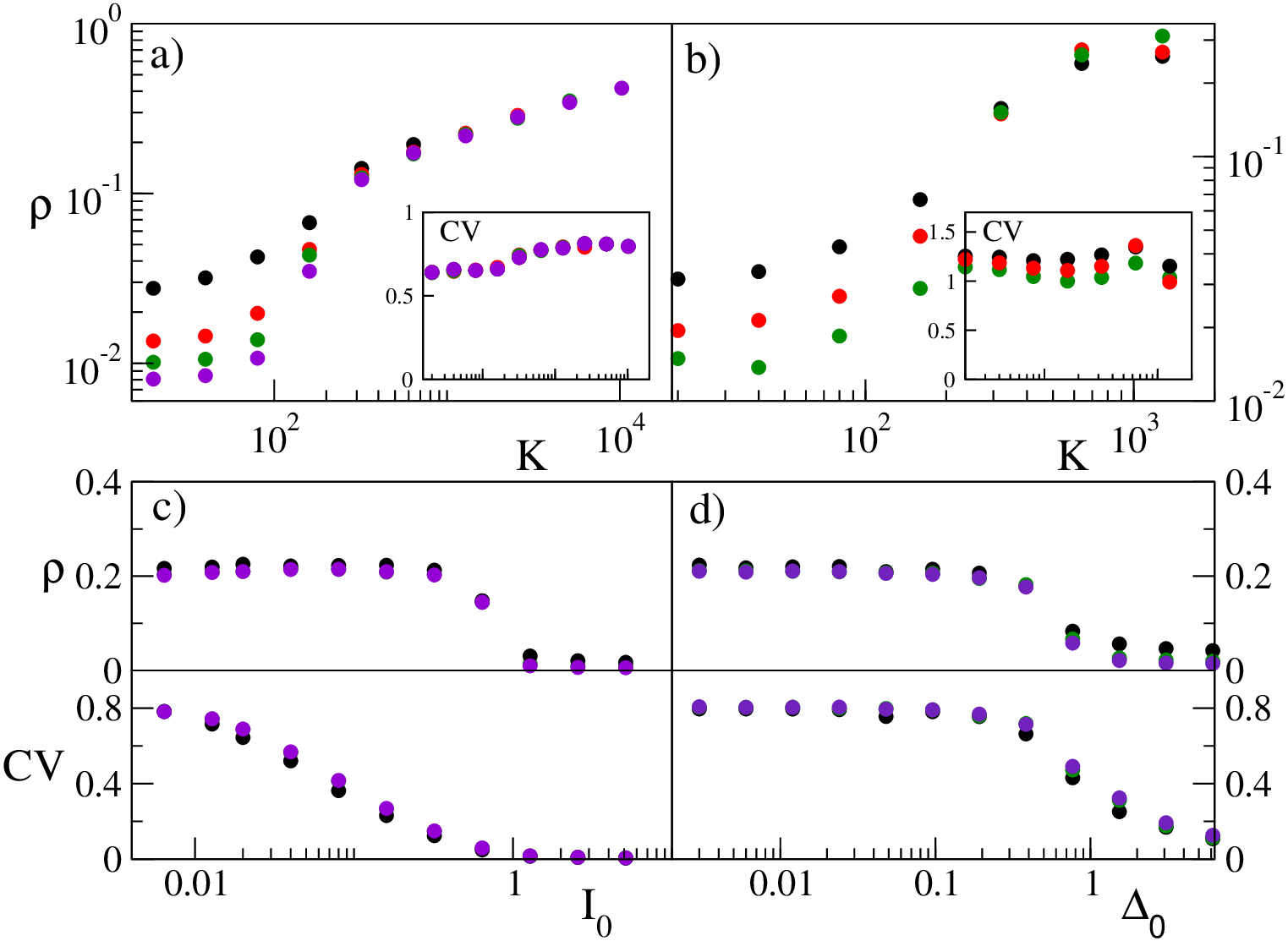}
\end{centering}
\caption{\label{fig:2} Upper panels: order parameter $\rho$ versus $K$ for QIF (a) and ML (b),
the inset report the corresponding CVs. The lower panels display in the upper part
$\rho$ and in the lower one the $CV$ versus $I_0$ (c) and $\Delta_0$ (d) for the QIF.
The data refer to various system sizes: namely $N=2000$ (black), 5000 (red),
10,000 (green) and 20,000 (violet). The employed parameters are 
$I_0 = 0.1$, $g_0=5$, and $\Delta_0=1$ for ML (b); for QIF
$g_0=1$, $\Delta_0=0.1$, $I_0=0.006$, $K=1000$.
}
\end{figure}

Asynchronous dynamics is indeed observable also for our models
for sufficiently sparse networks (small $K$), indeed 
a clear transition is observable from an asynchronous state to collective
oscillations for $K$ larger than a critical value $K_c$.
As observable from Figs. \ref{fig:2} (a,b), where we report
the coherence indicator $\rho$ as a function of $K$ for various system sizes from
$N=2,000$ to $N=20,000$. In particular, $\rho$ vanishes as $N^{-1/2}$
for $K < K_c$ (as we have verified), while it stays finite above the transition 
thus indicating the presence of collective motion. 
This transition resembles those reported for sparse
LIF networks with finite synaptic time scales in \cite{golomb2000,luccioli2012}
or with finite time delay in \cite{brunel1999,brunel2000}.
However, Poissonian-like dynamics of the single neurons has
been reported only in \cite{brunel1999,brunel2000}.
 
In the present case, in both the observed dynamical regimes the microscopic dynamics remains quite irregular for all the considered
$K$ and system size $N$, as testified by the fact that ${CV} \simeq 0.8$ for the QIF
and ${CV} \geq 1$  for the ML  (as shown in the insets of Fig. \ref{fig:2} (a,b)).
The relevance of the microscopic fluctuations for the existence of
the collective oscillations in this system can be appreciated by
considering the behaviour of $\rho$ and $CV$ as a function of the
external current $I_0$ and of the parameter controlling the structural
heterogeneity, namely $\Delta_0$. The results of these analysis
are shown in Figs. \ref{fig:2} (c) and (d) for the QIF
and for $N=2,000$, 10,000 and 20,000. In both cases we fixed
a in-degree $K > K_c$ in order to observe collective oscillations
and then we increased $I_0$ or $\Delta_0$. 
\red{In both cases we observe that for large $I_0$ ($\Delta_0$)
the microscopic dynamics is now imbalanced with few neurons firing
regularly with high rates and the majority of neurons suppressed by this 
high activity. This induces a vanishing of the $CV$, 
which somehow measures the degree of irregularity in the microscopic dynamics.}
At large $I_0$ the dynamics of the network is controlled by neurons definitely supra-threshold and the dynamics becomes mean-driven \cite{renart2007,angulo2017}.
The same occurs by increasing $\Delta_0$, when the heterogeneity in the in-degree
distribution becomes sufficiently large only few neurons, the
ones with in-degrees in proximity of the mean $K$, can balance their activity,
while for the remaining neurons it is no more possible to satisfy
the balance conditions, as recently shown in  \cite{landau2016,pyle2016,farzad2017}.
As a result, COs disappear as soon as the microscopic fluctuations, 
due to the balanced irregular spiking activity, vanish.

{\it Effective Mean-Field Model.} \red{In order to understand the origin of these macroscopic oscillations we consider an exact macroscopic model recently derived
in \cite{montbrio2015} for fully coupled networks of pulse-coupled QIF 
with synaptic couplings randomly distributed according to a Lorentzian.
The mean-field dynamics of this QIF network can be expressed
in terms of only two collective variables (namely, $V$ and $R$), as follows
~\cite{montbrio2015}: 
\begin{equation}
\dot{R} =  \frac{R}{\tau_m} \left(2 V + \frac{\Gamma}{\pi}\right) 
\enskip, \enskip 
\dot{V} = \frac{V^2 + I}{\tau_m} + R {\bar g} -(\pi R)^2 \tau_m 
\label{mf_0}
\end{equation}
where ${\bar g}$ is the median and $\Gamma$ the HWHM of the Lorentzian distribution of the synaptic couplings.}

Such formulation can be applied to the sparse network 
studied in this Letter, indeed the quenched disorder in the connectivity
can be rephrased in terms of a random synaptic coupling \cite{di2014heterogeneous}. Namely, each neuron $i$ is subject to an average inhibitory synaptic current of amplitude $g_0 k_i R/(\sqrt{K})$ proportional to its in-degree $k_i$.
Therefore we can consider the neurons as fully coupled, but with
random values of the coupling distributed \red{as a 
Lorentzian  of median ${\bar g} = - g_0 \sqrt{K}$ and HWHM $ \Gamma = g_0 \Delta_0$. 
The mean-field formulation \eqref{mf_0} takes now the expression:}
\begin{eqnarray}
& \tau_m \dot{R} =  R ( 2 V + \frac{g_0 \Delta_0}{\pi}) 
\label{mfR}
\\
& \tau_m \dot{V} = V^2 + \sqrt{K}(I_0 - \tau_m g_0 R ) -(\pi R \tau_m)^2 \enskip .
\label{mfV}
\end{eqnarray}
As we will verify in the following,  
this formulation represents a quite good approximation of the collective dynamics of 
our network. Therefore we can safely employ such effective mean-field model 
to interpret the observed phenomena and to obtain theoretical predictions
for the spiking network.

\begin{figure}
\begin{centering}
\includegraphics[width=0.4\textwidth,clip=true]{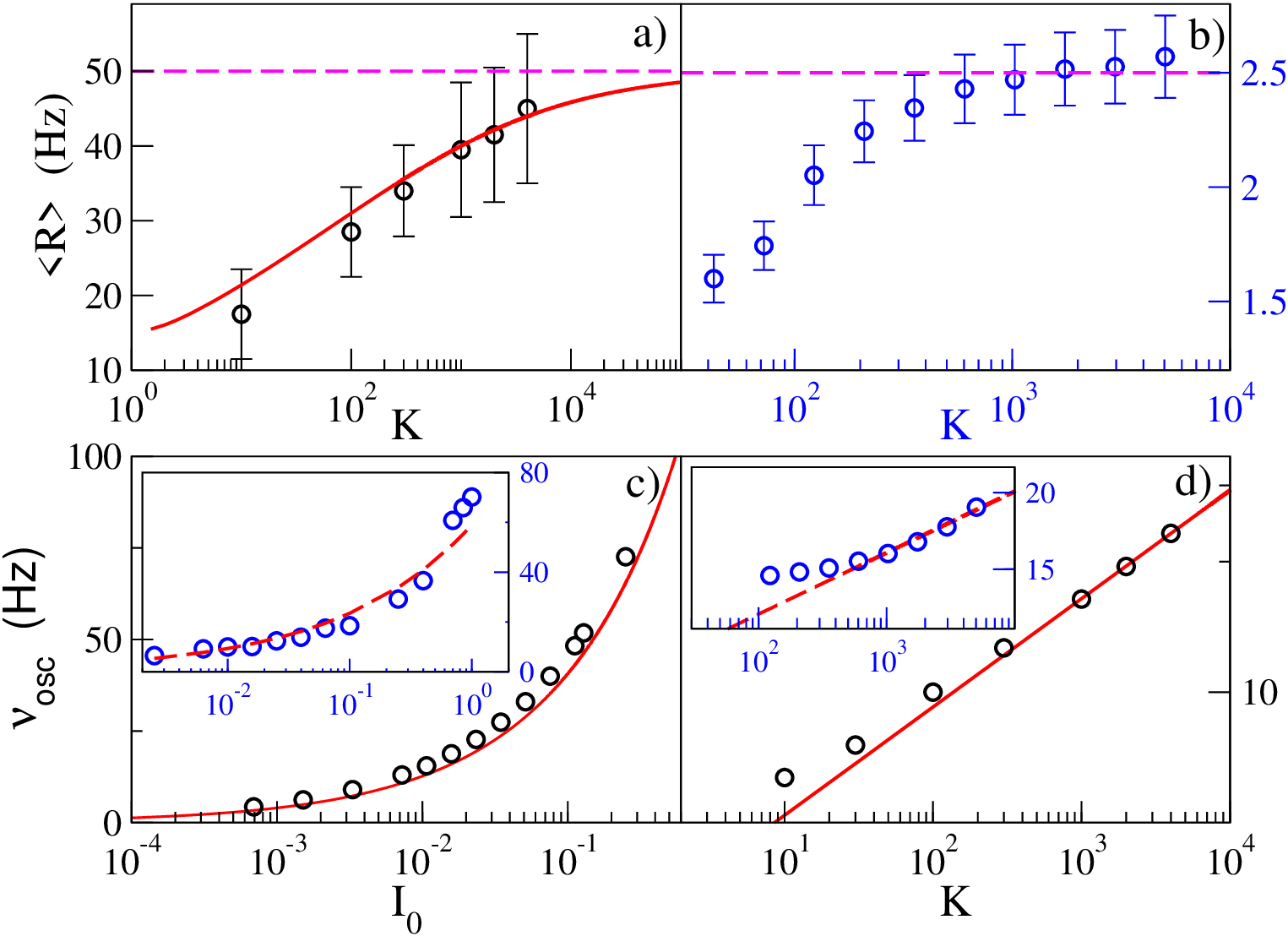}
\end{centering}
\caption{\label{fig:3}Upper panels: average firing rates $\langle R \rangle$ versus $K$ for QIF (a) and ML(b), the horizontal dashed (magenta) lines denote $R_a$ and the solid (red) line in (a) 
${\bar R}$ in Eq. \eqref{firing_teo}. The choice of parameters ($I_0,\Delta_0$) sets the dynamics as asynchronous: $(1,3)$ in (a) and $(0.05,8)$ in (b).
Lower panels: $\nu_{osc}$ versus $I_0$ (c) and versus $K$ (d) for the QIF, the insets display the same quantities for the ML. The red solid line in (c) refers to $\nu_{th}$, and in (d) to the theoretically predicted scaling $\nu_{th} \sim K^{\frac{1}{4}}$; the red dashed line in the inset of (c) and (d) to power-law fitting $\nu_{osc} \simeq I^{0.40}$ and $\nu_{osc} \simeq K^{0.10}$, respectively.
Oscillatory dynamics is observable for the selected parameter's values
($I_0,\Delta_0$): $(0.05,0.3)$ in (c) and $(0.05,0.5)$ in (d). Other parameters' values $N=10,000$, $g_0=1$, and $K=1000$ in (c). 
}
\end{figure}

Let us first consider the fixed point solutions $({\bar V},{\bar R})$ of Eqs. (\ref{mfR},\ref{mfV}).
The result for the average membrane potential is $ {\bar V} = (-g_0 \Delta_0)/(2 \pi)$,
while the firing rate is given by the following expression
\begin{equation}
{\bar R} \tau_m = \frac{g_0 \sqrt{K}}{2 \pi^2} \left(\sqrt{1+ \frac{4 \pi^2}{\sqrt{K}}\frac{I_0}{g_0^2}
+\frac{\Delta_0^2}{K}}-1\right) \enskip .
\label{firing_teo}
\end{equation}
This theoretical result reproduces quite well with the simulation findings for the QIF spiking network in the asynchronous regime (observable for sufficiently high $ \Delta_0$ and $I_0$) over a quite broad range of connectivities  (namely, $10 \le K \le 10^4$), as shown in Fig.\ref{fig:3} (a). At the leading order in $K$, the firing rate \eqref{firing_teo} is given by $R_a \tau_m = I_0/g_0$, which represents the asymptotic result to which the balanced
inhibitory dynamics converges for sufficiently large in-degrees irrespectively
of the considered neuronal model, as shown in Fig.\ref{fig:3} (a) and (b) for the QIF and ML 
models and as previously reported in  \cite{montefortePRX} for Leaky Integrate-and-Fire (LIF) neurons.
In particular, for the  ML model the asymptotic result $R_a$ is attained already
for $K \ge 500$, while for the QIF model in-degrees larger than $10^4$ are required.
 
The linear stability analysis of the solution $({\bar V},{\bar R})$ 
reveals that this is always a stable focus, characterized by two 
complex conjugates eigenvalues with a negative
real part $\Lambda_R \tau_m= -\Delta_0/2\pi$ and an imaginary
part $\Lambda_I \tau_m = \sqrt{2{\bar R} \tau_m(2\pi^2 {\bar R} \tau_m+\sqrt{K}g_0)-(\Delta_0/2\pi)^2}$.
\red{The frequency of the relaxation oscillations towards the stable fixed point solution
is given by $\nu_{th}=\Lambda_I/2\pi$. This represents a good approximation of the frequency $\nu_{osc}$ of the
sustained collective oscillations  observed in the QIF network}
over a  wide range of values ranging from ultra-slow rhythms to 
high $\gamma$ band oscillations, as shown in Fig. \ref{fig:3} (c).
Furthermore, it can be shown that $\nu_{th}$ predicts the correcting scaling of $\nu_{osc}$ for the QIF for sufficiently large DC currents and/or median in-degree $K$, namely $\nu_{th} \approx I_0^{1/2} K^{1/4}$ (as shown in Fig. \ref{fig:3} (c) and (d)). For the ML we observe similar scaling behaviours for $\nu_{osc}$, with
slightly different exponent, namely $\nu_{osc} \approx I_0^{0.4}$ and
$\nu_{osc} \simeq K^{0.10}$, however in this case we have not a theoretical prediction to compare with (see the insets of Fig. \ref{fig:3} (c) and (d)).

\begin{figure}
\begin{centering}
\includegraphics[width=0.5\textwidth,clip=true]{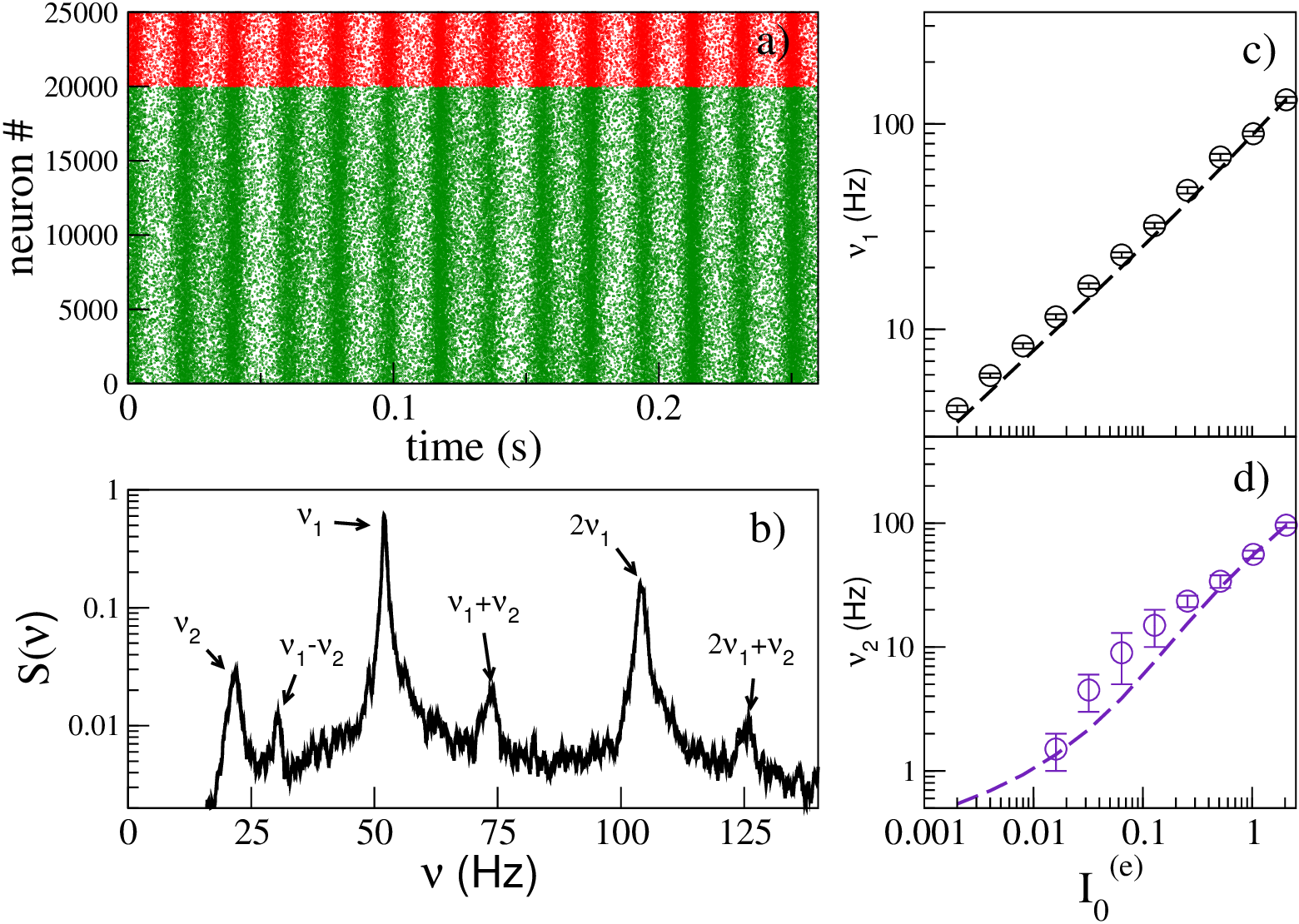}
\end{centering}
\caption{\label{fig:4} \red{The raster plot for a network of $N_E = 20,000$ excitatory (green) and $N_I = 5,000$ inhibitory (red) QIF neurons is displayed in (a). In (c) and (d) the COs' frequencies, measured from the power spectrum  $S(\nu)$ of the mean voltage $V(t)$ (shown in (b)), are reported as symbols versus the excitatory DC current $I_0^{e}$. The dashed lines are the theoretical mean-field predictions. The values of the parameters are reported in \cite{model}.}}
\end{figure}

{\it Excitatory-inhibitory balanced populations} So far we have
considered only balanced inhibitory networks, but in the cortex
the balance occurs among excitatory and inhibitory populations.
 To verify if also in this case collective
oscillations could be identified we have considered a neural
network composed of $80\%$ excitatory QIF neurons and
$20\%$ inhibitory ones (for more details on the considered model see the SM in \cite{model}). 
The analysis reveals that also in this
case collective oscillations can be observed in the balanced network
in presence of irregular microscopic dynamics of the neurons.
\red{This is evident from the
raster plot reported in Fig. \ref{fig:4} (a).
An important novelty is that now the oscillations are characterized by 
two fundamental frequencies as it becomes evident from the analysis of the power spectrum  $S(\nu)$ of the mean voltage $V(t)$
shown in Fig. \ref{fig:4} (b). As expected for a noisy quasi-periodic dynamics,
the spectrum reveals peaks of finite width at frequencies that can be obtained as linear combinations
of two fundamental frequencies $\nu_1$ and $\nu_2$. The origin of the noisy contribution
can be ascribed to the microscopic irregular firings of the neurons.}
Analogously to the inhibitory case, a theoretical prediction for the collective oscillation 
frequencies can be obtained by considering an effective mean-field model for the excitatory and
inhibitory populations of QIF neurons. 
The model is now characterized by 4 variables, i.e. the mean membrane
potential and the firing rate for each population, and also in this case
one can find as stationary solutions of the model a stable focus. However, 
the stability of the focus is now controlled by two couples of complex conjugate eigenvalues,
thus the relaxation dynamics of the mean-field towards the fixed point is quasi-periodic
(see the SM for more details \cite{model}).
A comparison between the theoretical values of these relaxation
frequencies and the measured oscillation frequencies $\nu_1$ and $\nu_2$
associated to the spiking network dynamics is reported in Figs. \ref{fig:4} (c) and (d)
for a wide range of DC currents, revealing an overall good agreement.
Thus suggesting that the mechanism responsible for the collective oscillations
remains the same identified for the inhibitory network.

{\it Conclusions.} We have shown that in balanced spiking networks 
with instantaneous synapses COs can be triggered by microscopic irregular fluctuations, whenever the neurons will share a sufficient number of common inputs.
\red{Therefore, for a sufficiently large in-degree the erratic spiking emissions can 
promote coherent dynamics. We have verified that the inclusion of a small synaptic 
time scale does not alter the overall scenario~\cite{tobe}.}

\red{It is known that heuristic firing-rate models, characterized
by a single scalar variable (e.g. the Wilson-Cowan model \cite{wilson1972}),
are unable to reproduce synchronization phenomena 
observed in spiking networks \cite{schaffer2013,devalle2017}. 
}
In this Letter, we confirm that the inclusion of the membrane 
dynamics in the mean-field formulation is essential to correctly
predict the frequencies of the COs, not only for finite synaptic 
times (as shown in~\cite{devalle2017}), but also for instantaneous synapses in dynamically balanced sparse networks. In this latter case, the internal time scale of the mean-field model controls the COs' frequencies over a wide and continuous range. \red{As we have verified, sustained oscillations can be triggered in the mean-field model by adding noise to the membrane dynamics. Therefore, an improvement of the mean-field theory here presented should include fluctuations around the mean values. A possible strategy could follow the approach reported in \cite{schaffer2013} to derive high dimensional firing-rate models from
the associated Fokker-Planck description of the neural dynamics  
 \cite{brunel1999,brunel2000}. Of particular interest would be to understand
if a two dimensional rate equation \cite{schaffer2013}  is
sufficient to faithfully reproduce collective phenomena also in balanced
networks.}

\red{Our results pave the way for a possible extension of the reported mean-field
model to spatially extended balanced networks \cite{kriener2013,rosenbaum2014,pyle2017,senk2018}
by following the approach employed to develop 
neural fields from neural mass models \cite{Coombes2006}.}


\begin{acknowledgments}
The authors acknowledge N. Brunel for extremely useful comments on a preliminary version of this Letter, as well as V. Hakim, E. Montbri\'o, L. Shimansky-Geier for enlightening discussions. AT has received partial support by the Excellence Initiative I-Site Paris Seine (No ANR-16-IDEX-008) and by the Labex MME-DII (No ANR-11-LBX-0023-01). The work has been mainly realized at the Max Planck Institute for the Physics of Complex Systems (Dresden, Germany) as part of the activity of the Advanced Study Group 2016/17 ``From Microscopic to Collective Dynamics in Neural Circuits”. 
\end{acknowledgments}

  

\begin{thebibliography}{43}
\bibitem[{CV()}]{CV}
\bibinfo{note}{The coefficient of variation $cv(i)$ for the neuron $i$ is the
  ratio between the standard deviation and the mean of the inter-spike
  intervals associated to the train of spikes emitted by the considered neuron.
  The average coefficient of variation is defined as $CV = \sum_i cv(i)/N$}.
\end{thebibliography}

\end{document}